\documentclass[a4paper,fleqn,usenatbib]{mnras}

\usepackage{newtxtext,newtxmath}

\usepackage[T1]{fontenc}
\usepackage{ae,aecompl}

\usepackage{graphicx}	
\usepackage{amsmath}	
\usepackage{amssymb}	
\usepackage{multirow}
\usepackage{float}
\usepackage{soulutf8}

\title{A new statistical method for characterizing the atmospheres of extrasolar planets}

\author[C. S. Henderson et al.]{
Cassandra S. Henderson$^{1}$, 
Andrew J. Skemer$^{1}$, 
\newauthor{Caroline V. Morley$^{2}$, Jonathan J. Fortney$^{1}$} 
\\
$^{1}$University of California, Santa Cruz
$^{2}$Harvard University}

\date{Accepted 2017 June 12. Received 2017 June 5; in original form 2017 May 11}

\pubyear{2017}

\begin{document}
\label{firstpage}
\pagerange{\pageref{firstpage}--\pageref{lastpage}}
\maketitle

\begin{abstract}
By detecting light from extrasolar planets, we can measure their compositions and bulk physical properties. The technologies used to make these measurements are still in their infancy, and a lack of self-consistency suggests that previous observations have underestimated their systemic errors. We demonstrate a statistical method, newly applied to exoplanet characterization, which uses a Bayesian formalism to account for underestimated errorbars. We use this method to compare photometry of a substellar companion, GJ 758b, with custom atmospheric models. Our method produces a probability distribution of atmospheric model parameters including temperature, gravity, cloud model (f$_{\rm sed}$), and chemical abundance for GJ 758b. This distribution is less sensitive to highly variant data, and appropriately reflects a greater uncertainty on parameter fits.  
\end{abstract}

\begin{keywords}
{methods: statistical, planets and satellites: atmospheres}
\end{keywords}



\section{Introduction} \label{sec:intro}
By comparing the measured spectral energy distributions of extrasolar planets to atmospheric models, it is possible to constrain exoplanet bulk physical properties such as temperature, cloudiness, and metallicity.  Exoplanet atmospheric characterization necessarily pushes instruments beyond their design capabilities, so systematic effects often dominate over atmospheric signatures. Underestimated errors and erroneous measurements have lead to incorrect inferences of exoplanet properties \citep{Burrows}.  Evidence of underestimated errors exist.  For example, \citet{Hansen} note discrepancies in multi-epoch \textit{Spitzer} photometry of transiting exoplanets.  Yet more measurements at greater resolution are needed for accurate characterization \citep{Croll, Lupu, Line}. A rigorous method to evaluate and recognize erroneous measurements is necessary to provide accurate interpretations of exoplanet properties.

In this paper, we study the properties of a direcly-imaged companion, GJ 758b, using a Bayesian technique that simultaneously evaluates the correctness of individual photometry points while fitting the data to a suite of atmospheric models.  GJ 758b is a T dwarf companion separated by 30 AU from a Solar type star with an age of 5-9 Gyr \citep{Thalmann, Janson}.  This object has been characterized before, however, previous work has demonstrated that its photometry is difficult to fit with atmospheric models or spectra of analog objects \citep{Vigan,Nilsson}.

In Section \ref{sec:data} we provide literature photometry data, information on the models we use, and a basic model fit. In Section \ref{sec:discussion} we explain why the photometry on GJ 758b gives us reason to believe that some of the data is erroneous.  In Section \ref{sec:pk} we describe our Bayesian approach to evaluate the quality of each data point, which is based on similar work to estimate the Hubble constant in the context of discrepant data \citep{Press}.  We compare GJ 758b's derived atmospheric properties with and without the Bayesian approach.

\begin{table} 
	\centering
	\caption{References: (1) - \citet{Vigan} (2)- \citet{Janson}, recalibrated for 15.5 pc}
	\begin{tabular}{ |l|l|c|l| } \hline 
		Filter & $\lambda (\mu m$) & Absolute Magnitude & Ref. \\ \hline
		Y2 & 1.022 & 19.19$\pm$0.20 & 1 \\ \hline
		Y3 & 1.076 & 18.43$\pm$0.10 & 1 \\ \hline
		J2 & 1.190 & 19.06$\pm$0.25 & 1 \\ \hline
		J & 1.250 & 17.60$\pm$0.20 & 2 \\ \hline
		J3 & 1.273 & 16.83$\pm$0.18 & 1 \\ \hline
		CH4s & 1.580 & 17.76$\pm$0.12 & 2 \\ \hline
		H2 & 1.593 & 16.59$\pm$0.12 & 1 \\ \hline
		H & 1.650 & 18.18$\pm$0.20 & 2 \\ \hline
		H3 & 1.667 & 18.88$\pm$0.42 & 1 \\ \hline
		CH4l & 1.690 & $\geq$18.68 & 2 \\ \hline
		K$_\text{c}$ & 2.098 & 17.14$\pm$0.20 & 2 \\ \hline
		K1 & 2.110 & 17.03$\pm$0.21 & 1 \\ \hline
		K2 & 2.251 & 17.78$\pm$0.35 & 1 \\ \hline
		L' & 3.776 & 15.0$\pm$0.1 & 2 \\ \hline
		M$_\text{s}$ & 4.670 & $\geq$13.1 & 2 \\ \hline
	\end{tabular} \\
	\vspace{2mm}
	\label{tab:Photometry on GJ 758b}
\end{table}
 
\section{Data and Model Fitting} \label{sec:data}
\subsection{Photometry} \label{sec:photometry}
We compiled literature data on GJ 758b from three sources: the VLT planet finder, SPHERE \citep{Vigan}, Gemini North using NIRI, and Keck observatories using NIRC2 \citep{Janson}. \citet{Vigan} provide absolute magnitudes calibrated for a distance of 15.76 pc, while \citet{Janson} provide absolute magnitudes calibrated for 15.5 pc, which we recalibrated for 15.76 pc to maintain consistency with \citet{Vigan}. All absolute magnitudes are tabulated in Table \ref{tab:Photometry on GJ 758b}. The distance change accounts for a difference of 0.02 magnitudes, which in the cases of L', and Ms, does not change the magnitude within the known significant figures. We used filter curves from the VLT, Gemini North, and Keck websites - and used cryogenically measured filter transmission curves when available to calculate effective wavelengths and model photometry for section \ref{sec:model_fitting}.

~\\

\subsection{Models} 
We calculate model spectra assuming radiative--convective and chemical equilibrium, and including sulfide and salt condensates as described in \citet{Morley12, Morley14}. Various updates to the chemistry and opacities will be described in detail in an upcoming paper (Marley et al. in prep.) Briefly, the methane line lists have been updated using \citet{Yurchenko14} and the alkali line lists have been updated to use the results from \citet{Allard05}. Chemical equilibrium calculations are based upon previous thermochemical models \citep{Lodders02, Visscher06, Visscher12}, and have been revised and extended to include higher metallicities. 

The model grid parameters we vary include temperature, gravity, metallicity, and cloud sedimentation parameter f$_{\rm sed}$ (see \citet{Ackerman01}). Our model grid contains 660 models, comprising temperatures of 450, 475, 500, 525, 550, 575, 600, 625, 650, 675, 700 K, surface gravities, of 30, 100, 300 and 1000 m/s$^2$, metallicities of [M/H]=0, 0.5, and 1.0, and cloud sedimentation efficiency of f$_{\rm sed}$=1, 2, 3, 5 and cloud-free. 

We linearly interpolate within this grid in order to better sample the error distribution of our parameters, increasing our effective grid size from 660 models to 17000 interpolated models. We have shown in \citet{Skemer16} that this interpolation minimally effects the results for similar ranges of parameters.

\subsection{Model Fitting}\label{sec:model_fitting}
To determine how well a model fits, we estimate model absolute magnitudes using the model SEDs and the filter transmission curves described in Section \ref{sec:photometry}. We compare the model magnitudes to the measured magnitudes in Table \ref{tab:Photometry on GJ 758b} to analyze the goodness of fit per model.

We used a Vega model from \citet{Rieke} which we smoothed by 0.3\%, as recommended, before using it to generate zero point fluxes.

We assume a radius for GJ 758b of 1.05 $R_J$ based on gas-giant planet evolution models which are well constrained to within 5\% by the age, mass, and separation of the system \hbox{\citep{Fortney}}. When not using a prior on the radius, it tended towards unphysically small radii  without significantly improving the fit.

We discarded models that were inconsistent with 3$\sigma$ upperlimits in the CH4l and M$_s$ filters (and otherwise we did not include these upperlimits in our new statistical method in Section \ref{sec:pk}).

\begin{figure*}
	\centering
	\includegraphics[width=19cm, height=10cm]{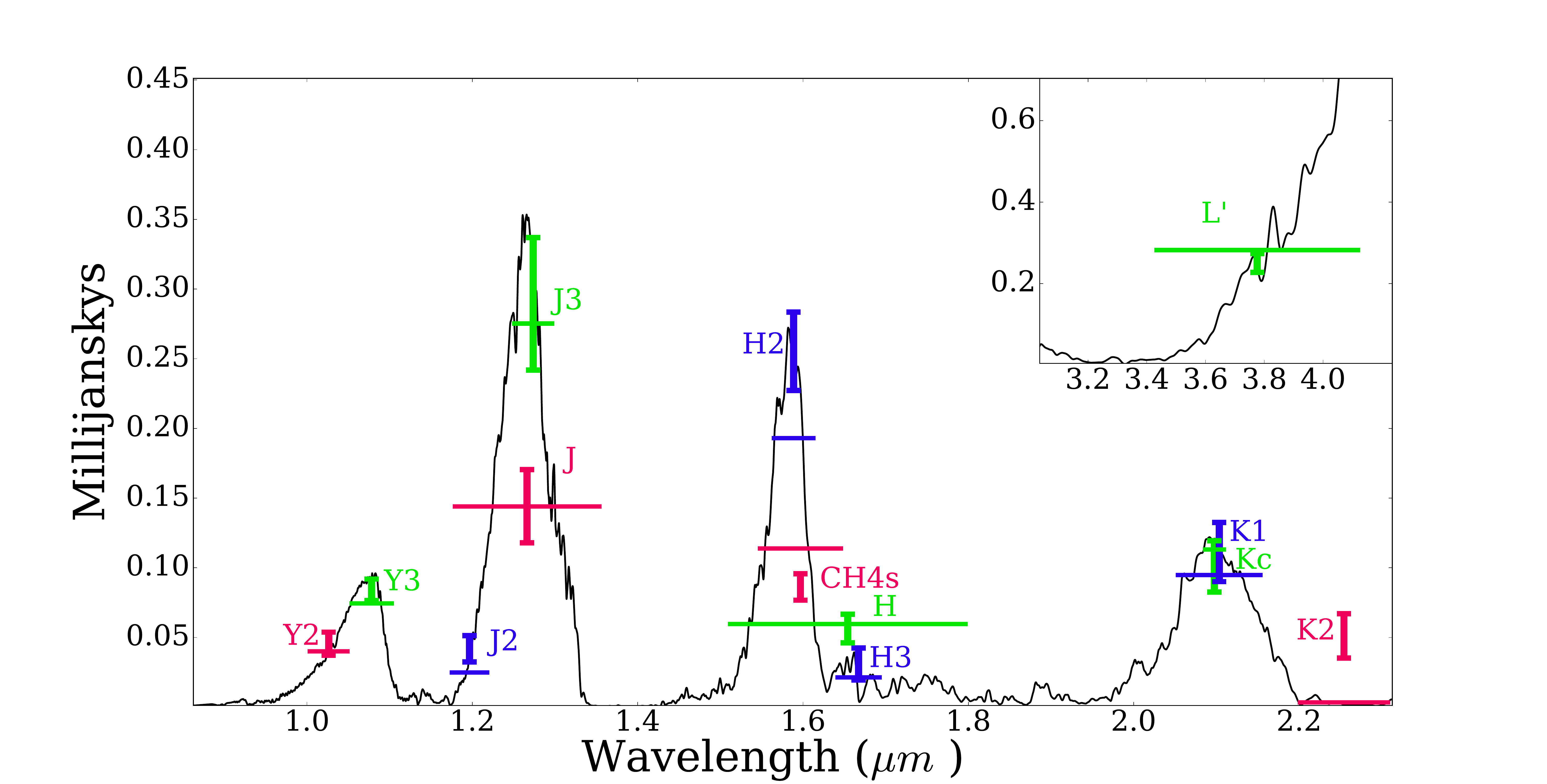}
	\caption{Best fit atmospheric model, with parameters $T=600K, G=250m/s^2, f_{\rm sed}=4.5, [M/H] = 0.0$. Color is added for contrast. The height of the horizontal bars is the expected flux integrated over the filter transmission curve, and the width is the filter width. }
	\label{fig:bestfit}
\end{figure*}

\begin{figure*}
	\centering
	\includegraphics[width=19cm, clip]{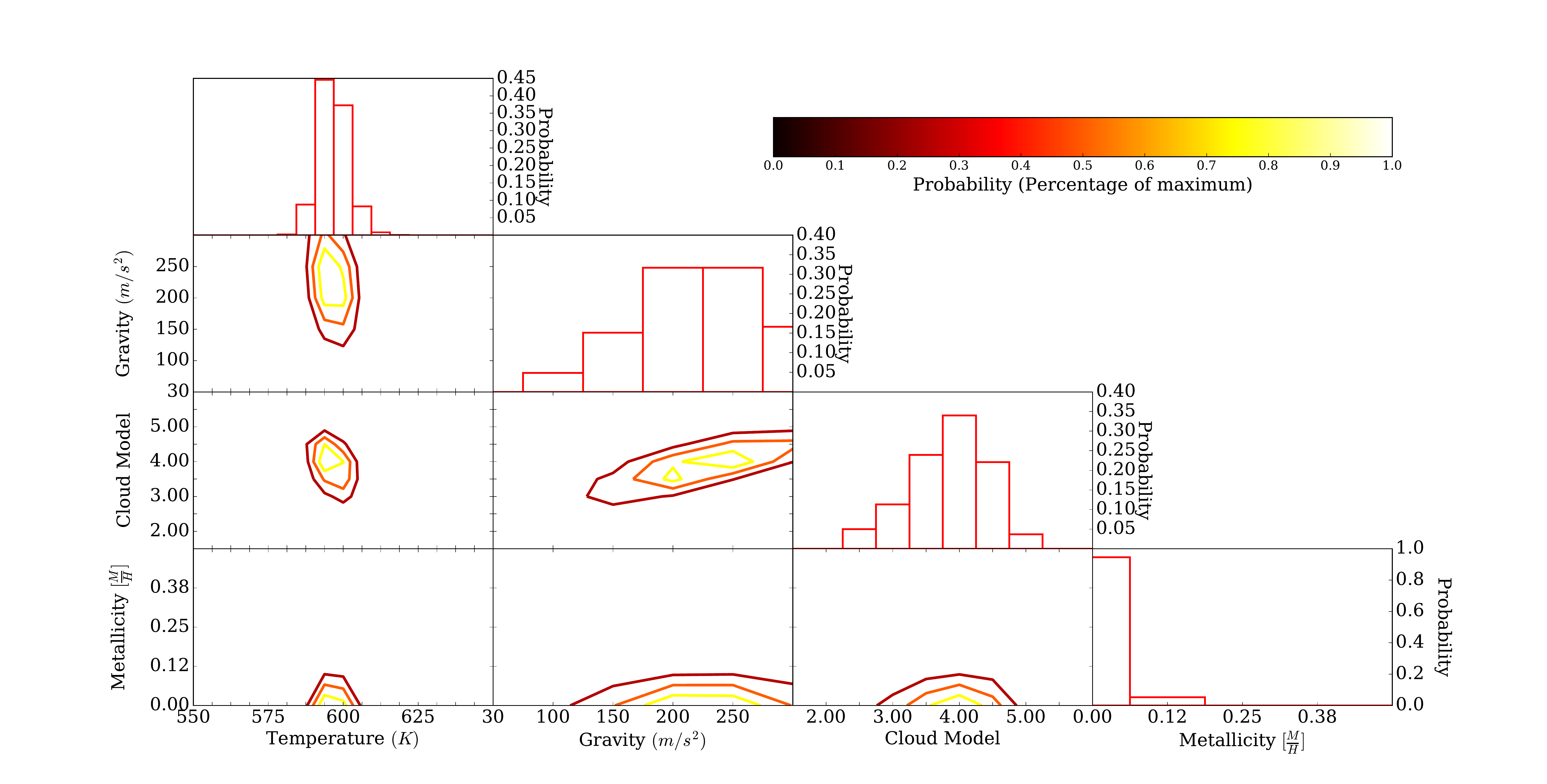}
	\caption{Standard PDF generated by comparing atmospheric models against given photometry. Contours are drawn at lines of 75, 50, and 25 percent of max probability.}
	\label{fig:corner_nopk}
\end{figure*}

Using a standard Bayesian fit for Gaussian errors ($P \propto \frac{1}{\sigma}e^{-\frac{{\chi}^2}{2}}$) normalized over all models, we found best fit parameters which we detail in Figure \ref{fig:corner_nopk}, and a best fit model shown in Figure \ref{fig:bestfit}. 

Notably, no model was able to simultaneously fit the CH4s and H2 photometry, which overlap in wavelength as shown in Figure \ref{fig:bestfit}. The best fit model reached $2.6\sigma$ for CH4s and $2.5\sigma$ for H2. Models which fit CH4s better fit H2 worse, and vice versa. Also, no model was able to fit K2: the best models had K2 magnitudes $> 8\sigma$ off. The best fit from the interpolated grid of models, had a reduced chi-square of 7.01.

\section{Bad models or bad data?} \label{sec:discussion}

\subsubsection{H2 and CH4s}

The CH4s and H2 filters overlap in wavelength, but report different flux measurements. We were not able to simultaneously fit the CH4s and H2 photometry within $2.5\sigma$ with self consistent atmospheric models, and a model that fits would require a very strange shape. 

\begin{figure}
	\centering
	\vspace{-5pt}
	\includegraphics[clip,width=.5\textwidth]{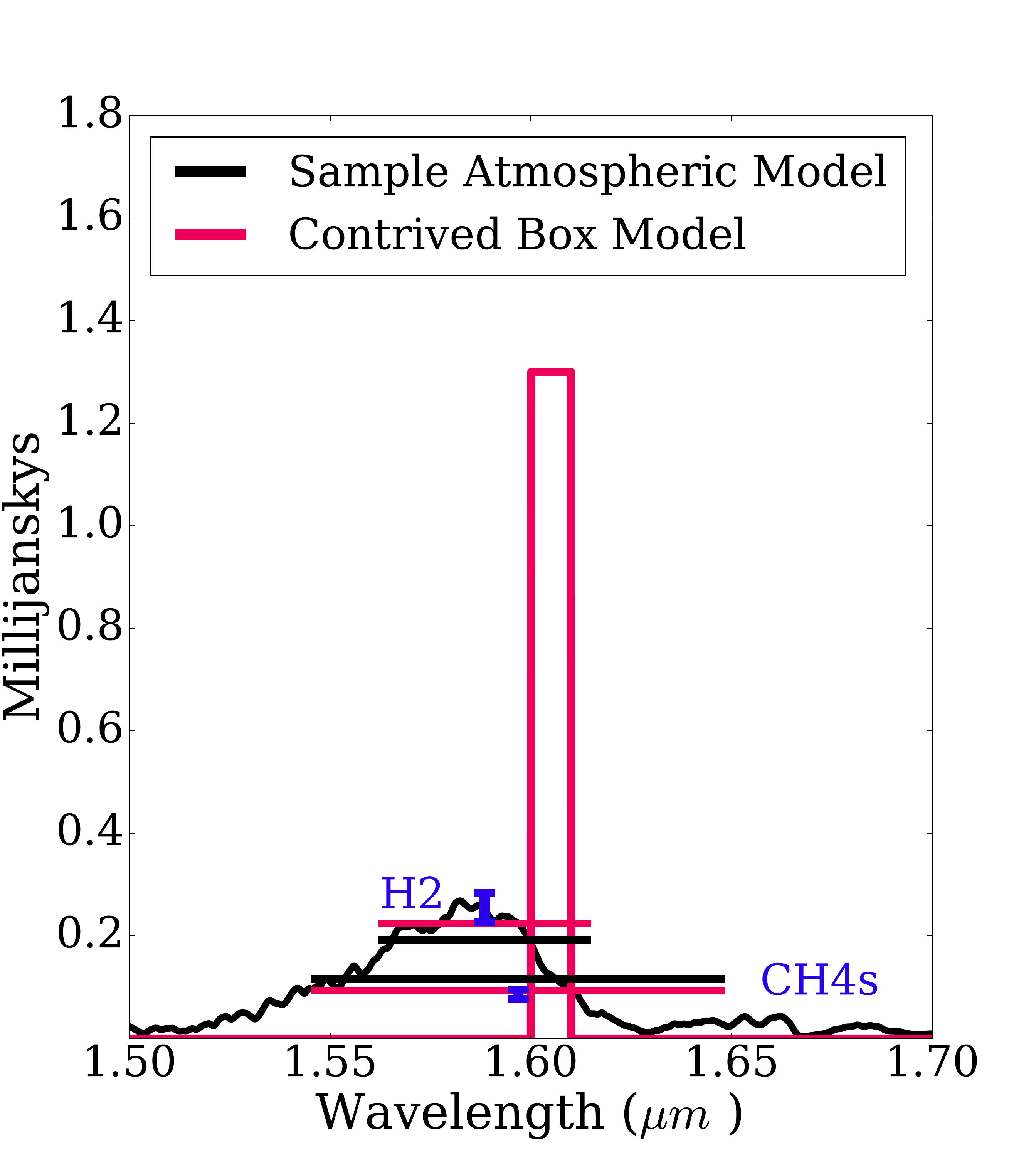}
	\caption{A synthetic model generated to fit CH4s and H2 simultaneously within $1\sigma$, which appears very unphysical compared to a realistic one, which falls at $2.7\sigma$ for CH4s and $2.6\sigma$ for H2.}
	\label{fig:synthetic_model}
\end{figure}

To demonstrate the problem, we constructed an arbitrary box model (free parameters: cut-on wavelength, cut-off wavelength and height) that best fits the CH4s and H2 photometry (see Figure \ref{fig:synthetic_model}).  This box could fit both points within 1 $\sigma$, due to its tuned shape, which sits on a low part of the H2 filter curve, and a high part of the CH4s filter curve.  However, the box model is far too discrepant from the self-consistent atmosphere models to be plausible.  The box model function has a reduced chi square of 0.89 and the atmosphere model, using only H2 and CH4s, has a reduced chi square of 6.88. The atmospheric model was chosen for the best simultaneous CH4s and H2 fit, with parameters $T=625K, g=300m/s^2, f_{\rm sed}=5.0, [M/H] = 0.0$ and a radius R=$0.85R_J$ chosen for best fit. Despite allowing the radius to vary to unphysically low amounts, we were still unable to find a model to fit both points simultaneously.

Since no realistic model could fit both H2 and CH4s, we conclude that these data points are inconsistent with themselves.

\subsubsection{Flux in K2 band}
We were unable to find any physical models that could fit K2 within 3$\sigma$. Flux is not expected at this wavelength \citep{Morley}. {The spectrum in the model spectra at the wavelengths of the K2 filter (~2.2-2.3 micron) is dominated by methane opacity. For the model spectra to be brighter in K2 to match the observed data, methane would need to be present at lower abundances than predicted. Disequilibrium carbon chemistry, which can decrease the methane abundance, is observed in brown dwarfs and exoplanets of similar temperatures (\hbox{\cite{Hubeny07}}; \hbox{\cite{Saumon06}}; \hbox{\cite{Konopacky13}}). However, for GJ 758b, decreasing the methane abundance to match the K2 photometry would also change other regions of the model spectrum. For example, the observed flux within the H3 filter (~1.6 micron), which is also dominated by methane opacity, would significantly increase. By this argument, we conclude that the the observed K2 photometry is not consistent with the rest of the spectrum, unless there is an unexpectedly substantial problem with the methane line list at near-infrared wavelengths. }

Assuming the K2 photometry is incorrect, including K2 in a statistical analysis will artificially constrain which models are preferred.  Removing this photometry point, or increasing its formal error, would increase the quality of our atmospheric fitting analysis.

K2 photometry could be correct and the models wrong, but the result of directly using highly variant photometry prevents productive model fitting of the rest of the data.

\section{Press and Kochanek's Method} \label{sec:pk}
To address the problem of data that is impossible to fit, we implemented Press and Kochanek's method \citep{Press}, which considers the possibility of underestimated errorbars.
\subsection{Background}
Press and Kochanek's method was originally used to determine the Hubble constant, using varied and contradictary data, where some select data points with small errors weighted the averaged value for the constant disproportionately compared to other measurements, despite being rather far from most other measurements. The method was introduced in order to address this problem rigorously, without discarding data entirely or arbitrarily. 

To do this, \cite{Press} defined a probability function where (1) individual data points have an unspecified probability of being ``correct'' or ``incorrect'', (2) data that are ``correct'' are assumed to have Gaussian-distributed errors as defined by the data's error bars, and (3) data that are ``incorrect'' are assumed to have Gaussian-distributed errors that are increased by some factor, which is defined at the outset of the analysis.  By marginalizing over the probability of each data point being ``correct'' or ``incorrect'', the technique is able to rigorously reject erroneous data.  A mathematical representation of this method is described in the appendix, Appendix \ref{sec:appendix}.

In \citep{Press}'s original implementation, the implicit assumption was that there is a single value of the Hubble constant that does not vary between measurements. { Discrepancies between measurements are then internally inconsistent.  When applying the method to SED fitting, the analogous assumption is that one of our model SEDs is correct and that the astrophysical source does not vary between measurements.  The assumption that one's model grid contains a ``correct'' model is implicit in all Bayesian model fitting, however we acknowledge that it is an additional assumption not considered by Press 1997.  In the case of GJ 758, the additional assumption is justified by multiple measurements at different wavelengths, which together, conflict with models so badly that they cannot all be correct.  Although our use of the method is different, the mathematical formalism is similar, as described in Appendix } \ref{sec:appendix}.

Overall, the method serves to desensitize the analysis to data that is hard to fit, which would ordinarily skew the result.


\subsection{Application to GJ 758b}
Given the difficulty of fitting CH4s and H2 simultaneously, as well as fitting K2, we consider the possibility that the errorbars for our data set from Table \ref{tab:Photometry on GJ 758b} are underestimated by a factor of 2 (note that increasing this fixed parameter has a negligable effect on the final results as shown below).  All photometry points are treated the same---for each photometry point, we assume, with a probability that is a free parameter, that the error bar might be underestimated by a factor of 2.  We evaluate the multi-dimensional posterior probability on a grid that includes the atmosphere model parameters, and the probability that each photometry point is correct or incorrect.

\begin{figure*}
	\centering
	\includegraphics[width=19cm, clip]{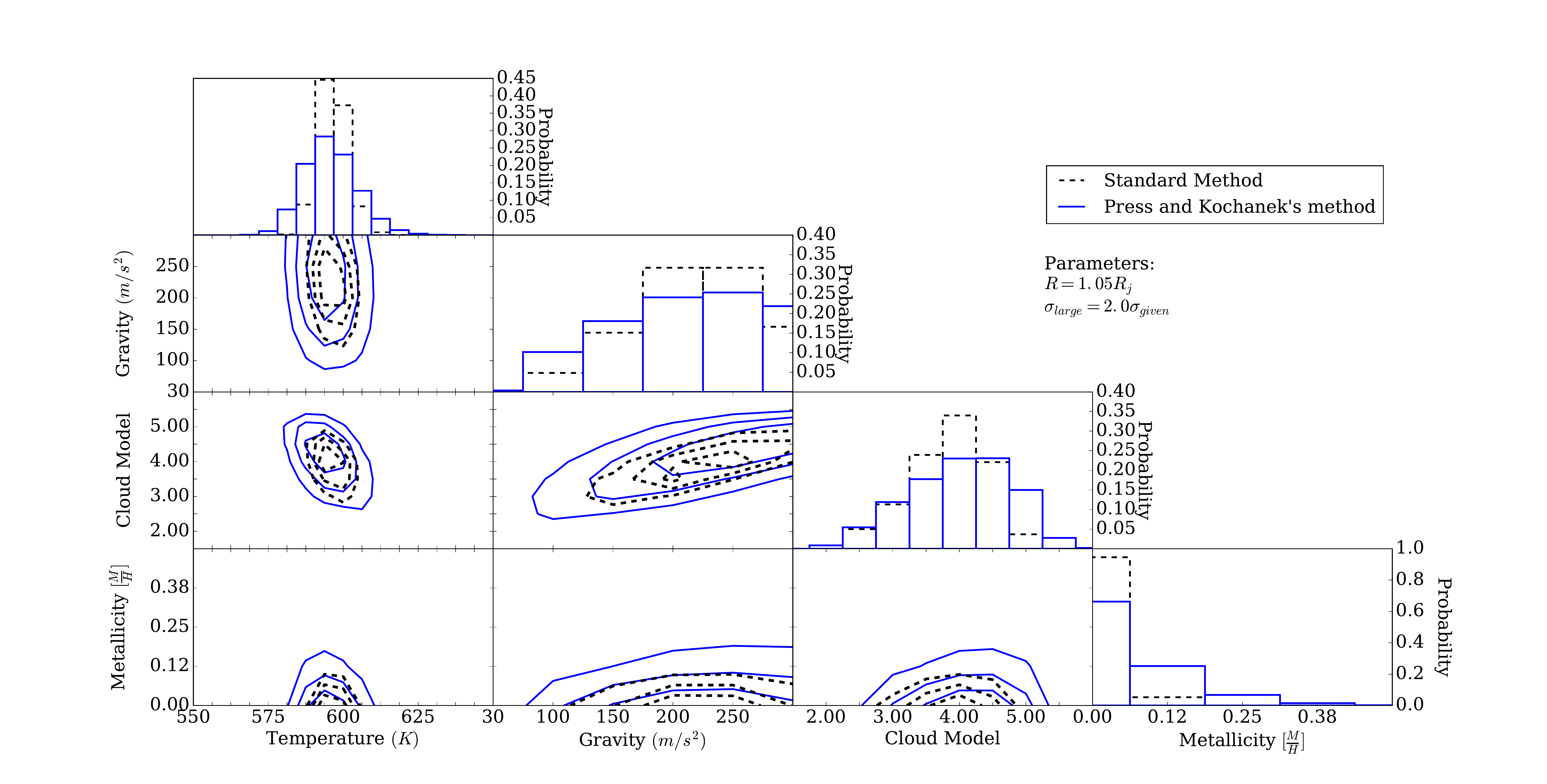}
	\caption{Probability distribution of parameters with and without the use of the statistical method. Note the increased width of probability distributions when using the method. Contours are at lines of 75, 50, and 25 percent of max probability. }
	\label{fig:cornerplot}
\end{figure*}

We plot the marginalized posterior probabilities of this method in Figure \ref{fig:cornerplot}. We note that the variance in the parameter PDF is larger than without the method, which suggests that we have removed the constraints of incorrect photometry. \cite{pstats} provides a probability of correctness per data point, which we plot in Figure \ref{fig:probgood}. This formula is detailed in Appendix \ref{sec:formula}, but in short it measures the consistency of each data point compared to the rest of the set.  The best fit model using this technique ended up being the same as the best fit model from Section \ref{sec:model_fitting} (shown in Figure \ref{fig:bestfit}).

\subsection{Method with prior}
\begin{figure*}
	\centering
	\vspace{-5pt}
	\includegraphics[width=17.5cm, clip]{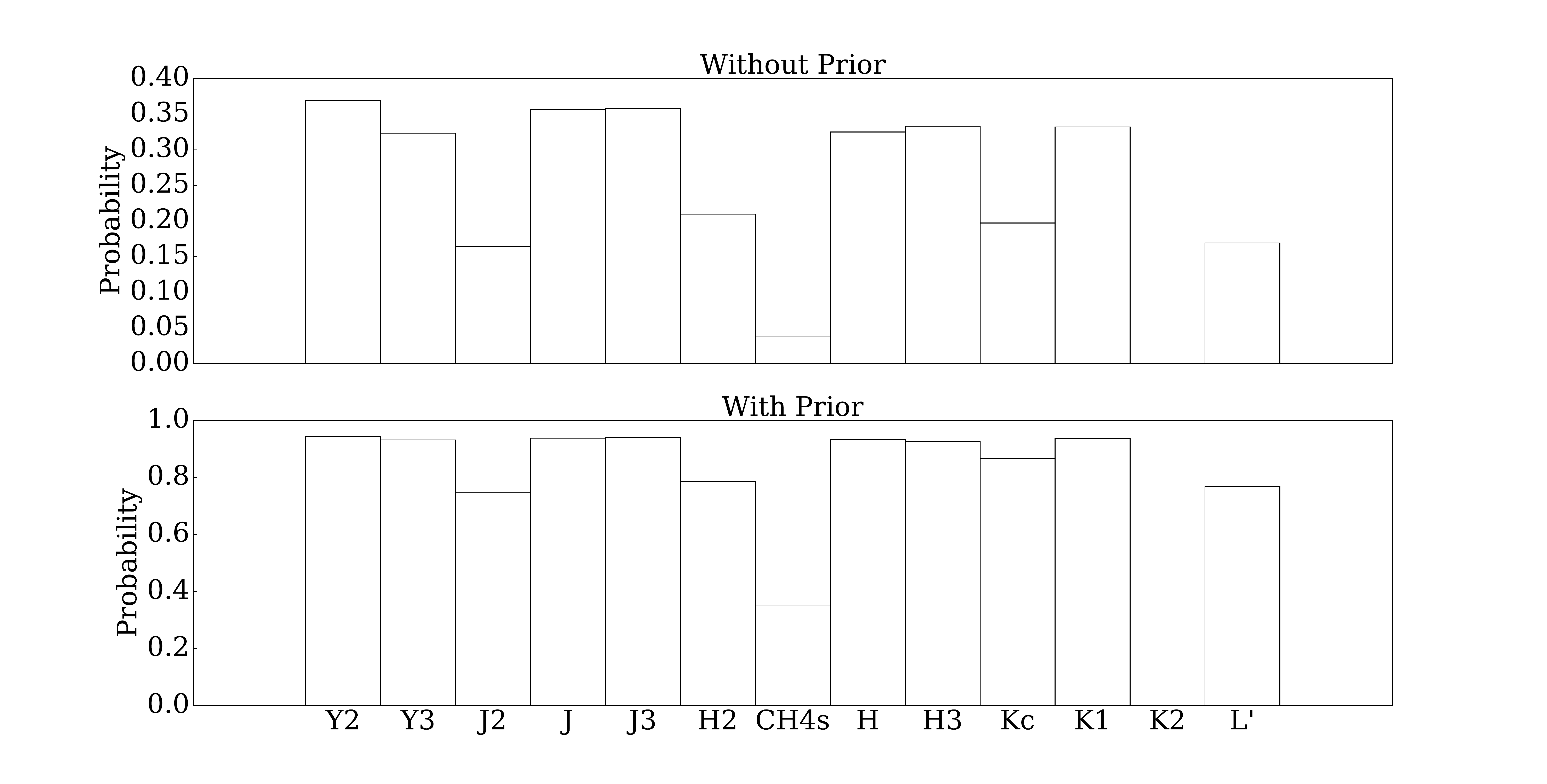}
	\caption{Press and Kochanek's provided formula for the correctness of any given point, which is really just a ratio between the probability with and without the use of the method, per point. These are plotted with and without an initial prior on the probability of point correctness of 0.9. In the instance without the prior, the total tends to lower probabilities, which is explained in the appendix, Appendix \ref{sec:appendix}.}
	\label{fig:probgood}
\end{figure*}

As shown in Figure \ref{fig:probgood} the method's estimate that each individual photometry point is correct is rather low: most photometry points have a $\sim$30\% chance of being correct, while the outlier points are assigned lower probabilities.  If we want to take the optimistic viewpoint that most observations result in correct error estimates, we can apply a prior that gives the observer the benefit of the doubt.  The prior fixes the probability that an individual photometry point is correct, which requires more substantial improvements to fits for the increased errorbar to become relevant.  The mathematical formalism is presented in Appendix \ref{sec:prior}.

\subsection{Method with varied parameters}
\begin{figure*}
	\centering
	\includegraphics[width=19cm, height=10cm]{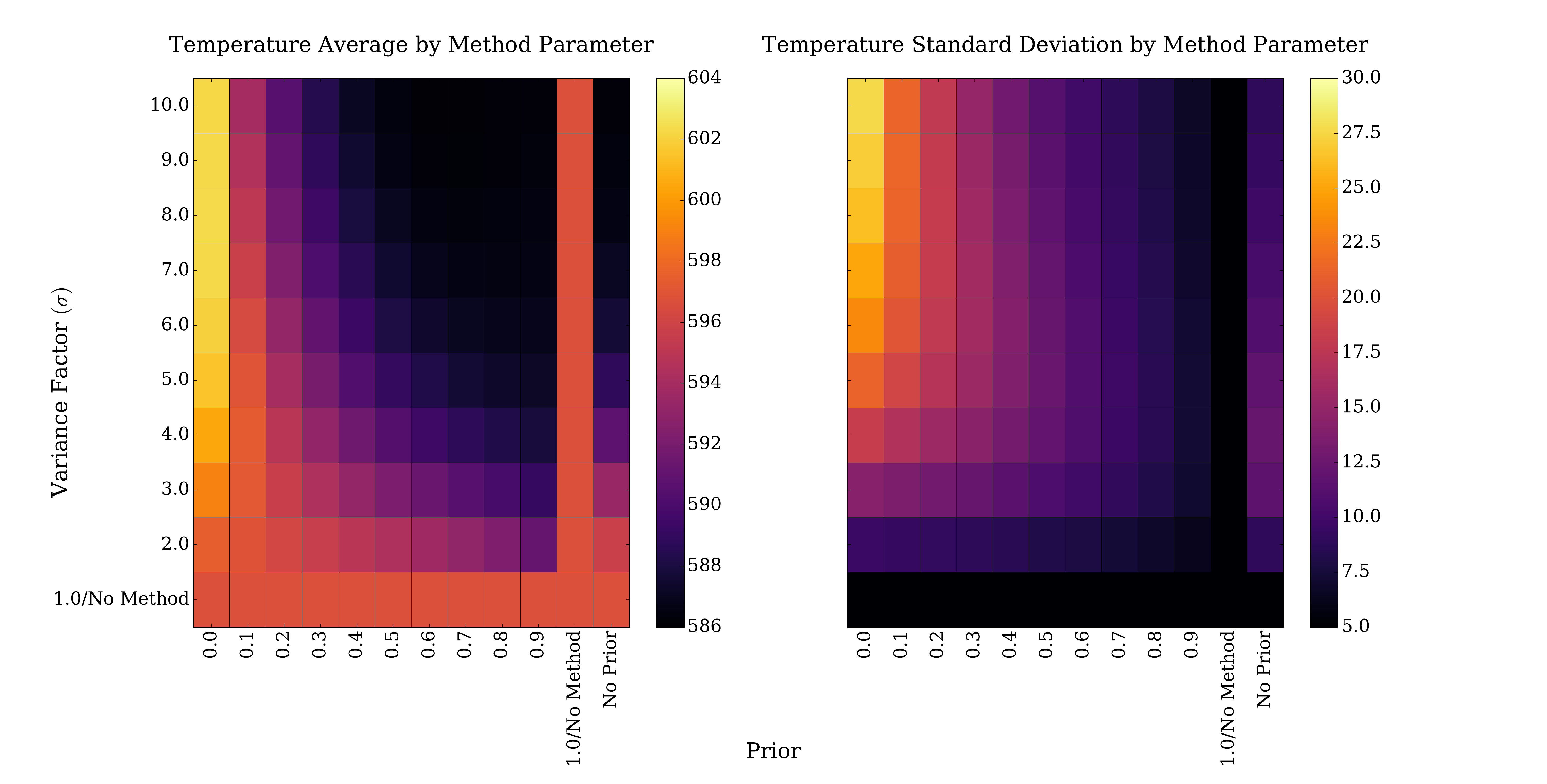}
	\caption{Temperature model parameter fit for varied parameters in our statistical method. For the temperature we found that even a small use of Press and Kochanek's method, with an optimistic prior and small variance factor, was notably different from the analysis without the method.}
	\label{fig:params}
\end{figure*}

We try using a variety of priors, and changing the factor which ``incorrect'' errors are increased by, to see how important method parameters are in changing the atmospheric parameter fit distributions. We plot our temperature fits and uncertainties in Figure \ref{fig:params} for different paremeters used in the method. Note that the least aggressive use of the method (assume that 90\% of the data are correct) still has a considerable effect on the temperature probability distribution.

\subsection{Results for GJ 758b}
The results of our analysis suggest that GJ 758 b's derived parameter errors increase when some photometry errors are assumed to be underestimated.  With a prior of 90\% correctness on the data (i.e. 1 or 2 incorrect data points) - the method decisively concludes that K2 has an underestimated errorbar, and that between CH4s and H2, CH4s is more likely to have an underestimated errorbar. 

The use of a prior on the correctness of each data point affects our results as shown in Figure \ref{fig:params}. In short the variance increases and the best fit changes marginally with any use of the method, but a more aggressive use (low prior or high measurement variance), within reason, does not have significantly different results from a less aggressive (high prior or low measurement variance) use of the method. 

The difference in parameter fit demonstrated in Figure \ref{fig:params}  indicates the change in temperature fit as a result of the method. The fit shifts up by 10K on the temperature when a larger (reasonable) variance factor is used. This varies from 1 to 2 $\sigma$ of the temperature fit, which may not be important for temperature but could dramatically effect the fit for other parameters, especially when an atmospheric retrieval method is used for generating models. Poor constraints from highly variant or inconsistent data ought to lead to poor fits on parameters, and we find that this statistical method can accomplish this for photometry with underestimated uncertainties.  

\subsection{Implications of Exoplanet Variability}

{We considered the possibility that GJ 758b exhibits variability that effectively causes an additional, unaccounted error term. \hbox{\cite{Radigan}} measure T-dwarf variabilities up to 1.6\% in the near-infrared due to rotation. However, this is far below the $\sim$10-20\% errors on photometric points used in our analysis.  For isolated objects, variability may contribute to ``under-estimated'' error bars.  For exoplanet imaging in the high-contrast regime, derived errors tend to be larger than the underlying variability \hbox{\citep{Apai}}.}

\section{Conclusion} \label{sec:conclusion}
To characterize exoplanet atmospheres, we must be able to fit them with atmospheric models. When we have reason to consider some of our errorbars to be underestimated, we can use Bayesian statistics to handle erroneous data quantitatively.

We implemented a statistical method to consider some of our errorbars as being larger than initially estimated on photometry for substellar companion GJ 758b, in order to address inconsistency in the data. Using our method consistently increased the standard deviation on the parameter fit of our atmospheric models, making it less sensitive to a small number of highly variant data points. \hbox{\cite{Hansen}} has suggested that a key to characterizing exoplanets is translating underestimated uncertainties to uncertainties on model parameter fits, and we find our result to be consistent with this goal. In the case of the photometry used in this analysis, our method also rigorously put into question our data points in the CH4s and K2 bands.

We conclude that the problem of characterizing exoplanet atmospheres with photometry that has underestimated errorbars can be addressed by increasing the errorbars rigorously using this method. We emphasize its use for low signal-to-noise photometry-based characterizations, including transit photometry, especially for new instruments such as \textit{James Webb Space Telescope}, where systemic error may not be immediately understood. We suggest that this method can also be used to make claims about the consistency of previous characterizations, as well as general trends of underestimated errorbars, as is necessitated by \cite{Burrows} and \cite{Hansen}.

\section{Acknowledgments}
We thank the referee for providing their very helpful report. We also thank the UC Santa Cruz \textit{LAMAT} program as well as the UC Santa Cruz Other Worlds Lab for their generous funding for this project.

\bibliographystyle{aasjournal}
\bibliography{bib}



\appendix

\section{Appendix: Press and Kochanek statistics}
\label{sec:appendix}
\subsection{Composite probability formula}
We've adapted Press and Kochanek's formula from \hbox{\cite{pstats}} for use in characterizing exoplanet atmospheres. As for regular model fitting, each measurement $X_i$ is compared to a model value $\theta_i$ which is computed using the filter curve used for $X_i$ and the model $\theta$ (as well as a Vega model). These values are compared using a Gaussian distribution and their given uncertainty, and the probability of the measurement being good is $\textrm{Prob}(X_i | \theta_i , \sigma_{\textrm{given}})$.
 
The product of all $\textrm{Prob}(X_i | \theta_i , \sigma_{\textrm{given}})$ gives a likelihood function for a $\theta$'s goodness, and this allows us to find the best fitting model. The probability of each model being correct given some measurements and errorbars follows in Equation \ref{eq:1}.

$\textrm{Prob}(\theta)$ represents the Bayesian normalization condition that one model must be correct. In this formula, $p$ iterates over the probability that the data are correct. This formula assumes no knowledge about the correctness of the observer. 

Because the probability functions are added, the larger errorbar term will dominate only if it significantly improves the fit. 

\subsection{Normalizing the probability function}
For calculating the probability of each data point being correct, we must normalize each data point to each model. The resultant normalization condition is Equation \ref{eq:2}.

Therefore each term in the product must be multiplied by some constant contained within $\textrm{Prob}(\theta)$ which is the $\frac{1}{n}th$ root of the unnormalized summed product, before integration. In this method, the only free parameter is the relation between $\sigma_{\textrm{given}}$ and $\sigma_{\textrm{large}}$.
\subsection{Adding a prior on p} \label{sec:prior}
We can add a prior on $p$ by instead evaluating the composite probability formula for p = prior. This leads to Equation \ref{eq:3} for the probability of a model being correct.

This results in the same normalization condition as before, except without the probability integration.
\subsection{Point correctness formula} \label{sec:formula}
Press and Kochanek's formula for the correctness of a given data point is adapted to Equation \ref{eq:4}.

Simply put, this is the ratio of goodness of fit when the original probability is used as opposed to when the composite probability is used for the $ith$ data point in particular. It will be higher when using a larger errorbar does not result in a better fit, and lower when it does. Therefore, the method weighs how well a data point fits, for each model, against the fit of every other data point in the set, to compose the probability that it is correct. This measures the consistency of a data point with all other data points, using the models as a metric.

\begin{figure*}
\centering
\begin{equation} \label{eq:1}
\textrm{Prob}(\theta | X_1, X_2, X_3... X_n) = \int_{p=0.0}^{p=1.0} \prod_{i=1}^{n} \big[[p\cdot \textrm{Prob}(X_i | \theta_i, \sigma_{\textrm{given}}) + (1-p) \cdot \textrm{Prob}(X_i | \theta_i, \sigma_{\textrm{large}})]\cdot \textrm{Prob}(\theta)\big]\,dp
\end{equation}

\begin{equation} \label{eq:2}
\int\limits_{\textrm{all models}}^{\theta}\int_{p=0.0}^{p=1.0} \prod_{i=1}^{n} \big[[p\cdot \textrm{Prob}(X_i | \theta_i, \sigma_{\textrm{given}}) + (1-p) \cdot \textrm{Prob}(X_i | \theta_i, \sigma_{\textrm{large}})]\cdot \textrm{Prob}(\theta)\big]\,dp\, d\theta = 1
\end{equation}

\begin{equation} \label{eq:3}
\textrm{Prob}(\theta | p, X_1, X_2, X_3... X_n) =  \prod_{i=1}^{n} \big[[p\cdot \textrm{Prob}(X_i | \theta_i, \sigma_{\textrm{given}}) + (1-p) \cdot \textrm{Prob}(X_i | \theta_i, \sigma_{\textrm{large}})]\cdot \textrm{Prob}(\theta)\big]
\end{equation}
 
\begin{equation} \label{eq:4}
\textrm{Prob}(\textrm{ith good}) =  \frac{\int\limits_{}^{\theta}\int_{p=0.0}^{p=1.0} p\cdot \textrm{Prob}(X_i | \theta_i, \sigma_{\textrm{given}}) \cdot \prod_{k=1, \ k \neq i}^{n} \big[[p\cdot \textrm{Prob}(X_k | \theta_k, \sigma_{\textrm{given}}) + (1-p) \cdot \textrm{Prob}(X_k | \theta_k, \sigma_{\textrm{large}})]\cdot \textrm{Prob}(\theta)\big]\,dp\, d\theta}
{\int\limits_{}^{\theta}\int_{p=0.0}^{p=1.0} \prod_{k=1}^{n} \big[[p\cdot \textrm{Prob}(X_k | \theta_k, \sigma_{\textrm{given}}) + (1-p) \cdot \textrm{Prob}(X_k | \theta_k, \sigma_{\textrm{large}})]\cdot \textrm{Prob}(\theta)\big]\,dp\, d\theta}$$
\end{equation}
\caption{Appendix formulae}
\end{figure*}


\bsp	
\label{lastpage}
\end{document}